\documentclass[twocolumn,aps,pra,shownopacs,superscriptaddress]{revtex4}
\usepackage{graphicx}% Include figure files
\usepackage{amssymb}
\usepackage{amsmath}
\usepackage{amsthm}
\usepackage{newtxtext,newtxmath}
\usepackage{dsfont}
\usepackage{wasysym}
\usepackage{color}
\usepackage{lineno}
\usepackage{array}
\usepackage{changes}
\usepackage{multirow}
\usepackage{makecell}
\usepackage{xcolor}
\usepackage[colorlinks=true,linkcolor=blue,urlcolor=blue,citecolor=blue,pdfauthor={ },pdftitle={},pdfsubject={ },pdfkeywords={ }]{hyperref}
\graphicspath{{Figures/}}

\begin{document}
	
	\title{Experimental realization of quantum Zeno dynamics for robust quantum metrology}
	
	\date{\today}
\author{Ran Liu}
\affiliation{Institute of Quantum Precision Measurement, State Key Laboratory of Radio Frequency Heterogeneous Integration, College of Physics and Optoelectronic Engineering, Shenzhen University, Shenzhen 518060, China}
\affiliation{Quantum Science Center of Guangdong-Hong Kong-Macao Greater Bay Area (Guangdong), Shenzhen 518045, China}
\author{Xiaodong Yang}
\email{yangxd@szu.edu.cn}
\affiliation{Institute of Quantum Precision Measurement, State Key Laboratory of Radio Frequency Heterogeneous Integration, College of Physics and Optoelectronic Engineering, Shenzhen University, Shenzhen 518060, China}
\affiliation{Quantum Science Center of Guangdong-Hong Kong-Macao Greater Bay Area (Guangdong), Shenzhen 518045, China}
\author{Xiang Lv}
\affiliation{College of Metrology Measurement and Instrument, China Jiliang University, Hangzhou, China}
\author{Xinyue Long}
\affiliation{Quantum Science Center of Guangdong-Hong Kong-Macao Greater Bay Area (Guangdong), Shenzhen 518045, China}
\author{Hongfeng Liu}
\affiliation{Department of Physics, State Key Laboratory of Quantum Functional Materials, and Guangdong Basic Research Center of Excellence for Quantum Science, Southern University of Science and Technology, Shenzhen 518055, China}
\author{Dawei Lu}
\email{ludw@sustech.edu.cn}
\affiliation{Department of Physics, State Key Laboratory of Quantum Functional Materials, and Guangdong Basic Research Center of Excellence for Quantum Science, Southern University of Science and Technology, Shenzhen 518055, China}
\affiliation{Quantum Science Center of Guangdong-Hong Kong-Macao Greater Bay Area (Guangdong), Shenzhen 518045, China}
\author{Ying Dong}
\email{yingdong@cjlu.edu.cn}
\affiliation{College of Metrology Measurement and Instrument, China Jiliang University, Hangzhou, China}

\author{Jun Li}
\email{lijunquantum@szu.edu.cn}
\affiliation{Institute of Quantum Precision Measurement, State Key Laboratory of Radio Frequency Heterogeneous Integration, College of Physics and Optoelectronic Engineering, Shenzhen University, Shenzhen 518060, China}
\affiliation{Quantum Science Center of Guangdong-Hong Kong-Macao Greater Bay Area (Guangdong), Shenzhen 518045, China}

\begin{abstract}
	
	Quantum Zeno dynamics (QZD), which restricts the system's evolution to a protected subspace, provides a promising approach for protecting quantum information from noise. Here, we explore a practical approach to harnessing QZD for robust quantum metrology. By introducing strong inter-particle interactions during the parameter encoding stage, we overcome the typical limitations of previous QZD studies, which have largely focused on single-particle systems and faced challenges where QZD could interfere with the encoding process. We experimentally validate the proposed scheme on a nuclear magnetic resonance platform, achieving near-optimal precision scaling under amplitude damping in both parallel and sequential settings. Numerical simulations further demonstrate the scalability of the approach and its compatibility with other control techniques for suppressing more general types of noise. These findings highlight QZD as a powerful strategy for noise-resilient quantum metrology.
	
\end{abstract}
\maketitle
\begin{figure}[b]
	\includegraphics[scale=0.48]{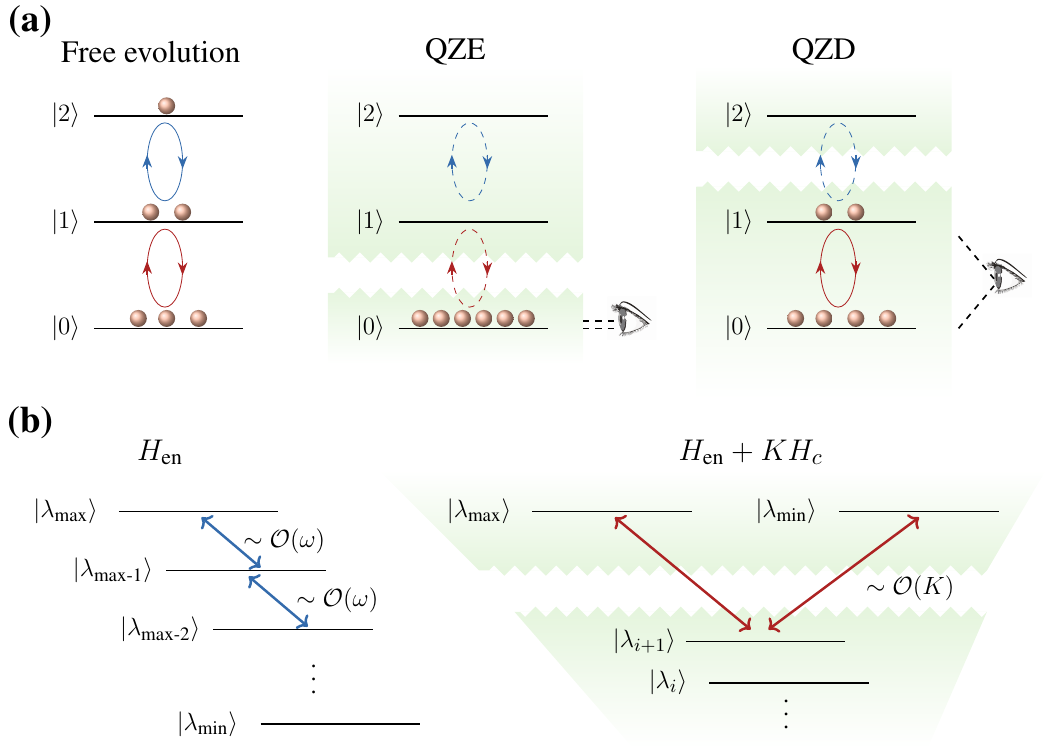} 
	\caption{Principle of QZD and its implementation Strategy.
	(a) Schematic diagram of the QZE and QZD. Solid (dashed) arrows indicate allowed (forbidden) transitions, while the eye symbols represent frequent measurements on the system. (b) Modification of energy-level structure induced by the strong coupling $KH_c$, where $|\lambda_\text{max}\rangle,|\lambda_\text{max-1}\rangle,\dots,|\lambda_\text{min}\rangle$ denote eigenvectors of $H_\text{en}$ corresponding to the maximal, second maximal, $\dots,$ and minimal eigenvalues, respectively. In the limit $K\to\infty$, the Zeno subspace $\mathcal H_\text{opt}=\text{span}\{|\lambda_\text{min}\rangle,|\lambda_\text{max}\rangle\}$ is dynamically isolated, and transitions outside $\mathcal{H}_\text{opt}$ are suppressed.}\label{Fig1}
\end{figure}

	\newpage
	\textit{Introduction.---}
	Quantum Zeno effect (QZE), which can freeze the evolution of a quantum state by exploiting frequent projective measurements \cite{Misra1977}, is a fundamental phenomenon in quantum mechanics. Since its first experimental observation in 1990 \cite{Itano1990}, QZE has attracted extensive theoretical \cite{kofman2000,PazSilva2012,slichter2016,Blumenthal2022,Zhang2023,Ronchi2024,Denton2024} and experimental interest \cite{Nagels1997,Streed2006,Bernu2008,Harrington2017,piacentini2017,Do2019,Virz2022}. One of the most intriguing extensions of QZE is the quantum Zeno dynamics (QZD).
	When frequently projecting onto a multidimensional subspace, the system governed by QZD can evolve away from its initial state while still remaining within the so-called Zeno subspace, as schematically illustrated in Fig.~\ref{Fig1}(a). This leads to the system's coherent evolution exhibiting intrinsic immunity to noise.
	Beyond projective measurements, QZD can also be induced by frequent unitary kicks or strong continuous coupling \cite{Facchi2004,burgarth2022}, which broadens its physical implications and enabling experimental realizations across diverse quantum platforms \cite{Raimond2010,Schafer2014,Signoles2014,bretheau2015}.

	Given its ability to protect coherent evolution from noise, QZD holds promising potential for robust quantum metrology. To bring this into reality, we face two primary challenges.
	First, quantum metrology can achieve unprecedented precision in parameter estimation through multi-particle entanglement \cite{Giovannetti2006,Giovannetti2011}, yet most QZD studies have only focused on single-particle systems. Implementing QZD in a multi-particle metrology framework with available resources remains a challenging problem. Second, the additional measurements, controls, or couplings required to realize QZD may introduce noncommutativity in the encoding dynamics. This potentially interferes with the parameter encoding and degrades the expected metrological precision \cite{Yuan2015,Li2024}. A critical issue is whether the noise-resisting power of QZD can be harnessed while ensuring its compatibility with the metrological process.
	
In this work, we propose a QZD-based robust quantum metrology scheme to resolve the above issues, and experimentally verify its feasibility on a nuclear magnetic resonance (NMR) platform. Specifically, QZD is realized through introducing a strong coupling between qubits during the encoding process. This is straightforward for systems like NMR with natural qubit interactions and easy to implement in programmable systems like superconducting circuits. Without requiring extra controls and frequent measurements, this strategy is thus favorable for implementing QZD in a multi-qubit system. Furthermore, we prove that the constructed QZD is intrinsically compatible with the encoding process. This indicates that the noise-resisting capability of QZD is preserved throughout the metrology process, and conversely, the expected parameter estimation precision remains unaffected by the presence of QZD.
 The proposed scheme is validated and benchmarked via the canonical quantum sensing scheme of  multi-qubit Ramsey interferometry experiment \cite{Degen2017, Pezz2018} on an NMR platform. 
 Experimental results demonstrate that our scheme restores the $1/N$ precision scaling under amplitude damping in parallel settings, and significantly enhances estimation precision by extending coherence time in sequential settings.
 Further numerical simulations  confirm the scalability of our scheme to larger systems and its potential for integration with dynamical decoupling techniques, highlighting its practical feasibility and advantages.

\begin{figure}
		\centering
			\includegraphics[scale=0.7]{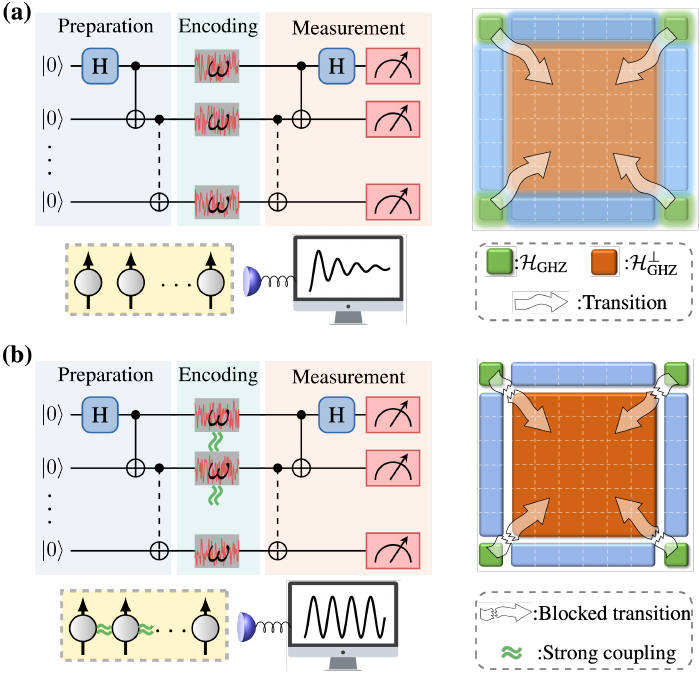} 
			\caption{Conventional and QZD-based  multi-qubit Ramsey interferometry. (a) and (b) show quantum circuits and schematic illustrations of the cases with and without QZD, respectively.
			 QZD is realized via strong inter-qubit coupling during the encoding process. Colored boxes indicate different subspaces in the system's Hilbert space.}\label{Fig2}
\end{figure}
	
	\textit{General framework of QZD.---} Consider a quantum system on a Hilbert space $\mathcal H$, evolving under $U(t)=e^{-iHt}$. Let ${P_n}$ be a complete set of orthogonal projectors satisfying $P_nP_m=\delta_{mn}P_n$ and $\sum_nP_n=\mathds 1$, which decompose the space as $\mathcal H=\oplus_n\mathcal H_{P_n}$ with $\mathcal H_{P_n}=\mathrm{Ran}P_n$. QZD can be implemented by inserting $N$ projective measurements at equal intervals, described by $V_N(t) = [P_nU(t/N)P_n]^N$. In the limit $N \to \infty$, the evolution reduces to
\begin{eqnarray}
\lim_{N\to\infty} V_N(t)=e^{-iH_Zt},
\end{eqnarray}
with effective Hamiltonian $H_Z = P_nHP_n$ acting within the Zeno subspace $\mathcal H_{P_n}$ \cite{Facchi2002}. If the initial state $\rho_0$ satisfies $P_n\rho_0P_n = \rho_0$, its evolution remains confined to $\mathcal H_{P_n}$.

Alternatively, QZD can be realized via strong continuous coupling. Let the coupling Hamiltonian be $KH_c$, with strength $K$, so that $U_K = e^{-i(H + KH_c)t}$. In the limit $K \to \infty$, a dynamical superselection rule emerges:
\begin{eqnarray}\label{strongcoup}
\lim_{K\to\infty}U_K=e^{-i(H_Z'+KH_c)},
\end{eqnarray}
where $H_Z' = \sum_n P_nHP_n$ and $H_c = \sum_n \eta_n P_n$ \cite{burgarth2022}. For $n\neq m$, $\eta_n\neq\eta_m$, indicating that the resulting Zeno subspaces are eigenspaces corresponding to distinct eigenvalues. Consequently, one can assign $H_c$ with specific degenerate levels to dynamically tailor the Hilbert space.

\textit{QZD for robust quantum metrology.---}We now outline the application of QZD to quantum metrology. A typical quantum metrology protocol includes: (1) preparing a probe state $|\psi(0)\rangle$; (2) encoding an unknown parameter $\omega$ via the Hamiltonian $H_\text{en}(\omega)$ over a duration $t$; and (3) measuring the final state $|\psi(t)\rangle$ to estimate $\omega$ \cite{Giovannetti2006,Giovannetti2011,huang2024}. The encoding Hamiltonian is typically taken as a multiplicative form $H_\text{en}(\omega) = \omega H_\text{en}$, where the optimal probe is $|\psi(0)\rangle = (|\lambda_{\min}\rangle + |\lambda_{\max}\rangle)/\sqrt{2}$, with $|\lambda_{\min/\max}\rangle$ being the eigenstates of $H_\text{en}$ corresponding to its minimal and maximal eigenvalues \cite{pang2017}. The ultimate precision is bounded by the quantum Cram{\'e}r-Rao bound (QCRB) $\Delta \omega \ge 1/\sqrt{\nu \mathcal F}$, with $\nu$ the number of independent measurements and $\mathcal F$ the quantum Fisher information (QFI) \cite{helstrom1969,Braunstein1994,petz2011}. In the absence of noise, the final state $|\psi(t)\rangle = \left( e^{-i\lambda_{\min} t} |\lambda_{\min}\rangle + e^{-i\lambda_{\max} t} |\lambda_{\max}\rangle \right)/\sqrt{2} $ evolves within the optimal subspace $\mathcal H_\text{opt} = \text{span}\{|\lambda_{\min}\rangle, |\lambda_{\max}\rangle\}$, leading to the maximal QFI $\mathcal F = (\lambda_{\max} - \lambda_{\min})^2 t^2$.

However, quantum metrology is often fragile to environmental noises. If there exist a noise term $\xi(t) H^{\prime}$ that does not commute with $H_\text{en}$, the system state will leak out of $\mathcal H_\text{opt}$, hereby degrading the precision. To suppress such leakage, we employ QZD by introducing a strong coupling $K H_c$ during encoding, where $\mathcal H_\text{opt}$ is required to be an eigenspace of $H_c$ with eigenprojector $P_\text{opt} = |\lambda_{\min}\rangle\langle \lambda_{\min}| + |\lambda_{\max}\rangle\langle \lambda_{\max}|$. That is, $|\lambda_{\min/\max}\rangle$ are degenerate eigenstates of $H_c$, distinct from others. Moreover, to ensure that QZD does not affect the metrological performance, a commutation condition $[H_c,H_\text{en}]=0$  must be satisfied  \cite{Yuan2015,Li2024}. This condition further reduces to $[P_\text{opt}H_cP_\text{opt},P_\text{opt}H_\text{en}P_\text{opt}]=0$ as the the probe state is prepared and evolves within $\mathcal H_\text{opt}$. Fortunately, since $P_\text{opt}$ is the eigenprojector of $H_c$, it follows that $P_\text{opt}H_cP_\text{opt}\propto\mathds 1_\text{opt}$ (the identity in $\mathcal H_\text{opt}$), inherently ensuring commutativity.

\begin{figure*}
		\centering
			\includegraphics[scale=0.89]{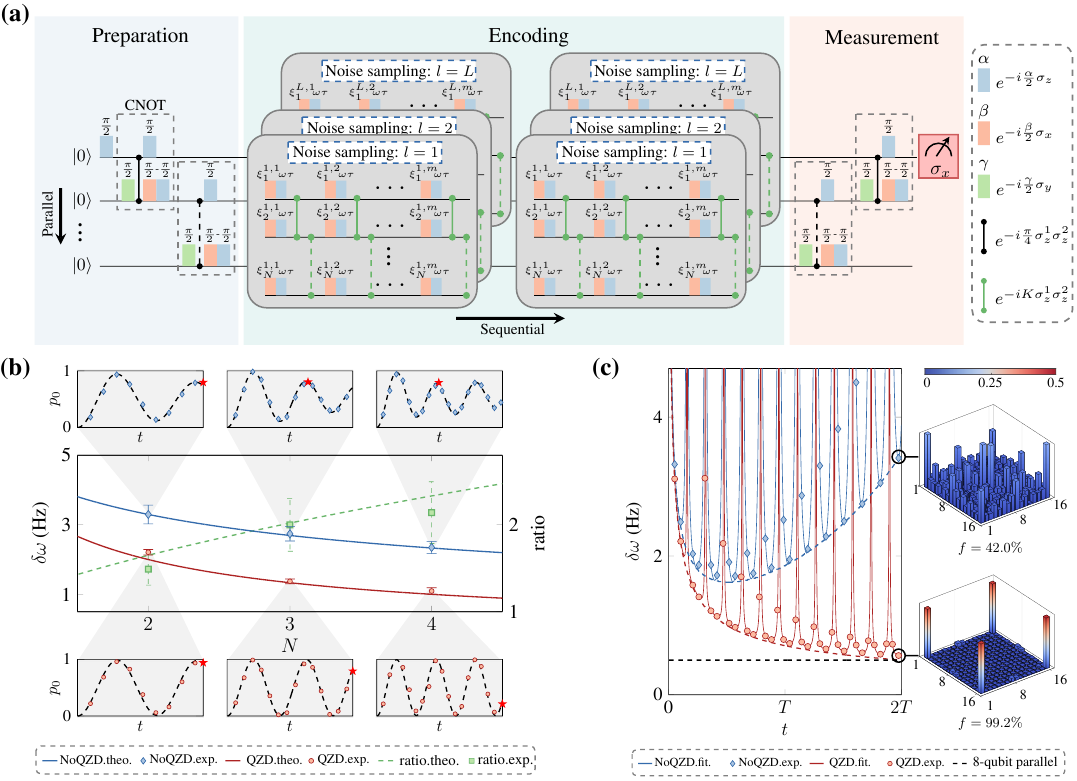} 
			\caption{Results of the QZD-based multi-qubit Ramsey experiments.
			(a) Explicit pulse sequence, where amplitude-damping noise is simulated via ensemble averaging, with $\xi_i^{L,m}$ denoting the amplitude of a transverse stochastic field. QZD is implemented through nearest-neighbor Ising interactions. 
			(b) Experimental results for the parallel sensing setting. The red star marks the optimal encoding time $t_\text{opt}$. With this encoding time fixed, the corresponding estimation precision $\delta \omega$ as a function of $N$ is shown in the main panel.
			 (c) Experimental results of a hybrid  parallel and sequential setting with $N=4$ and $t=2T$. The theoretical precision of the parallel setting with $N=8$ and $t=T$ (black dashed line) serves as a precision benchmark. The density matrices of the final states are reconstructed via quantum state tomography, where $f$ denotes their fidelity with encoded noise-free GHZ state.
			}\label{Fig3}
\end{figure*}

The effect of QZD in quantum metrology can be intuitively understood through the change of energy-level structure induced by the strong coupling $KH_c$. As shown in Fig.~\ref{Fig1}(b), the energy gaps associated with the encoding Hamiltonian $H_\text{en}$ are of order $\mathcal{O}(\omega)$. The introduction of $KH_c$ enlarges the energy separation between the encoding-relevant states $|\lambda_{\max/\min}\rangle$ and the rest of the spectrum to $\mathcal{O}(K)$. This energetic isolation suppresses noise-induced transitions out of the optimal subspace $\mathcal{H}_\text{opt}$. Moreover, due to the degeneracy of $|\lambda_{\max}\rangle$ and $|\lambda_{\min}\rangle$ under $H_c$, the added coupling does not contribute relative dynamical phases and thus leaves the metrological signal unaffected.

We now illustrate our scheme in a concrete setting---Ramsey interferometry using entangled multi-qubit probes. Consider estimating  a static field $\omega$ along the $z$-axis, described by $H_\text{en} = (\omega/2)\sum_i \sigma_z^i$. The optimal probe is the Greenberger-Horne-Zeilinger (GHZ) state $|\psi_\text{GHZ}\rangle = (|0\rangle^{\otimes N} + |1\rangle^{\otimes N})/\sqrt{2}$, which evolves within the optimal subspace $\mathcal{H}_\text{GHZ} = \text{span}\{|0\rangle^{\otimes N}, |1\rangle^{\otimes N}\}$.
  After a duration $t$, the state becomes $|\psi(t)\rangle = (|0\rangle^{\otimes N} + e^{iN\omega t}|1\rangle^{\otimes N})/\sqrt{2}$. Measurement in the GHZ basis yields transition probability $p_\text{GHZ} =[1 + \cos(N\omega t)]/2$. The estimation precision is bounded by $\Delta\omega \ge 1/\sqrt{\nu \mathcal{I}}$ \cite{helstrom1969,Braunstein1994}, with $\mathcal{I}$ the classical Fisher information (CFI). In the absence of noise, this saturates the QCRB and reaches the Heisenberg limit $\mathcal{F} = N^2t^2$. However, GHZ states are highly fragile and easily leave $\mathcal{H}_\text{GHZ}$ under noise \cite{Zhang2015,yang2020}, as illustrated in Fig.~\ref{Fig2}(a). Inspired by recent advances using many-body interactions to enhance metrological robustness \cite{Dooley2018,Zhou2020,Dooley2021,Chu2023,jiang2025}, we adopt accessible two-body couplings to realize QZD. We consider four typical nearest-neighbor interactions widely implemented across platforms: dipolar $H_c^\text{dip} = \sum_n (3\sigma_z^n \sigma_z^{n+1} - \vec{\sigma}^n \cdot \vec{\sigma}^{n+1})$, scalar (Heisenberg) $H_c^\text{sca} = \sum_n \vec{\sigma}^n \cdot \vec{\sigma}^{n+1}$, Ising $H_c^\text{Is} = \sum_n \sigma_z^n \sigma_z^{n+1}$, and quadratic Zeeman-type $H_c^\text{QZ} = (\sum_n \sigma_z^n)^2$. It is readily seen that all but $H_c^\text{sca}$ satisfy the criteria for constructing the Zeno subspace $\mathcal{H}_\text{GHZ}$ and thereby suppress noise leakage [Fig.~\ref{Fig2}(b)].

\textit{Experiment.---}The experiments are conducted at room temperature on a Bruker 300 MHz NMR spectrometer. The sample is the $^{13}$C-labeled trans-crotonic acid dissolved in $d_6$-acetone (see molecular structure and parameters in \cite{supp}). The four $^{13}$C nuclear spins serve as a four-qubit quantum sensor governed by the natural Hamiltonian
\begin{eqnarray}\label{naturalH}
H_{\mathrm{NMR}}=\sum_{i=1}^4 \pi \mu_i \sigma_z^i+\sum_{1 \leqslant i<j \leqslant 4} \frac{\pi}{2} J_{i j} \sigma_z^i \sigma_z^j.
\end{eqnarray}
Here, $\mu_i$ denote the chemical shifts and $J_{ij}$ represents the $J$-couplings. Starting from thermal equilibrium, the system is prepared into a pseudo-pure state (PPS) $\rho_{\mathrm{pps}} = [(1-\epsilon)/16] \mathds{1}^{4} + \epsilon|0\rangle\langle 0|^{\otimes 4}$ via spatial averaging, with $\epsilon \sim 10^{-5}$ denoting thermal polarization.

	\nocite{miller2012,taylor1997,Naghiloo2017,bylander2011,yan2013,Jin2012,Menke2022}
	
	The experimental procedure of the QZD-based metrology protocol proceeds as follows; see Fig.~\ref{Fig3}(a). Starting from the effective state $|0\rangle^{\otimes 4}$, a GHZ state $|\psi_\text{GHZ}\rangle$ is prepared via a Hadamard gate on qubit 1 followed by CNOT gates to the remaining qubits. The parameter $\omega$ is then encoded under $H_\text{en} = (\omega/2)\sum_i \sigma_z^i$ with amplitude-damping noise simulated via bath engineering \cite{soare2014e,Soare2014,Long2022}: a stochastic $x$-field $\xi_i^l(t)$ is applied in $L$ runs, each governed by $H_\text{AD}^l = \sum_i \xi_i^l(t) \sigma_x^i$ \cite{supp}. To implement QZD, we introduce an Ising coupling $K H_c^\text{Is}$ during encoding, which naturally arise in NMR spin systems.
	The encoding dynamics with and without QZD are thus described by the Hamiltonians $\tilde{H}_\text{en}^Z = H_\text{en} + H_\text{AD}^l + K H_c^\text{Is}$ and $\tilde{H}_\text{en} = H_\text{en} + H_\text{AD}^l$, respectively. As both are time-dependent and non-commuting, we discretize the encoding period into $M$ slices with fixed $\xi_i^{l,m}$ per slice. The dynamics are realized via Trotter-Suzuki decomposition \cite{suzuki1993}. Finally, the accumulated relative phase between $|0\rangle^{\otimes N}$ and $|1\rangle^{\otimes N}$ is extracted via multi-coherence measurement \cite{Laflamme2002} by applying the reverse GHZ circuit and measuring $\langle \sigma_x \rangle$ on the first qubit.

	We now present key parameters and techniques employed in our experiment. We set $\omega = 20$ Hz and discretize the total encoding time $T = 250$ ms into $M = 250$ slices. We find $L=10$ noise realizations are sufficient to accurately approximate the amplitude-damping noise \cite{supp}. 
	The engineered noise possesses a white-noise power spectral density, yielding the transition probability
\begin{eqnarray}\label{p_0}
p_0 = \frac{1}{2} \left[ 1 + \cos(N\omega t) e^{-N\Gamma t} \right],
\end{eqnarray}
with decoherence rate $\Gamma = 1.0$ Hz.  To suppress this noise via QZD, Eq.~\eqref{strongcoup} suggests that a sufficiently large coupling strength $K$ is required. According to the form of interaction in our NMR experiment, we employed the Ising coupling. Numerical simulations indicate that $K = 65$ is sufficient to mitigate the engineered noise, corresponding to a minimum $J$-coupling of $41.4$ Hz---well within the experimental range \cite{supp}. To achieve high-fidelity QZD, we compile the entire pulse sequence (see Fig.~\ref{Fig3}(a)) into a single shaped pulse using the gradient ascent pulse engineering (GRAPE) method \cite{khaneja2005}. Each pulse lasts 50 ms, divided into 1000 segments, with theoretical fidelities exceeding 99.5\% and robustness against inhomogeneities in the radio-frequency field.

	\begin{figure}
		\centering
			\includegraphics[scale=0.8]{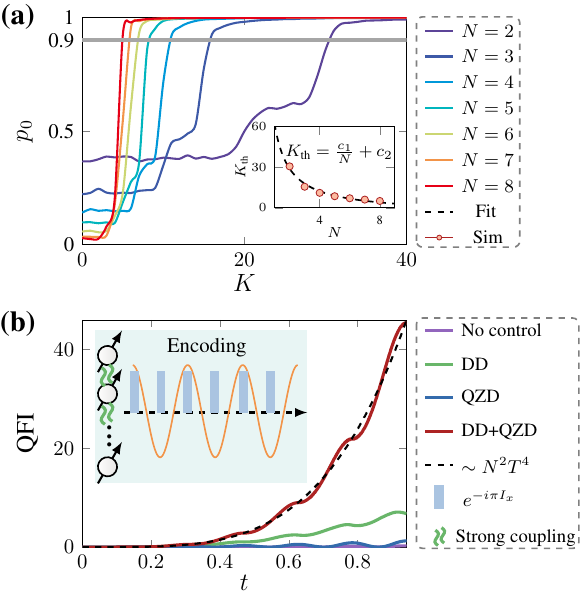} 
			\caption{Scalability of QZD-based quantum metrology scheme and its integration with dynamical decoupling.
(a) Ramsey fringe amplitude at $t = 2T$ versus coupling strength $K$ for various system sizes $N$ (colored lines). The gray line marks the threshold amplitude 0.9; intersections define the minimum coupling $K_\text{th}$. Inset: The scaling behavior of $K_\text{th}$ with $N$, which is fitted by $K_\text{th} = c_1/N+c_2$ with $c_1=66.4,c_2=-4.4$.
(b) QFI for ac field frequency estimation versus encoding time. QZD and DD are implemented via continuous strong coupling $K H_c^\text{Is}$ and periodic $\pi$-pulses along $x$, respectively (inset).}\label{Fig4}
\end{figure}

	\textit{Results.---}We first demonstrate our scheme in the parallel metrology setting, where sensing precision improves with increasing entangled qubit number \cite{omran2019,song2019}. Specifically, we perform 2-, 3-, and 4-qubit Ramsey interferometry experiments, sampling six points per period of $p_0$ during encoding. The frequency $\omega$ and decoherence rate $\Gamma$ are estimated by fitting $p_0$ to Eq.~\eqref{p_0}. As shown in the subfigures of Fig.~\ref{Fig3}(b), without QZD, $p_0$ showcases notable decays within the encoding duration $T$, yielding the fitted mean values of $\omega=$19.5 Hz and $\Gamma=$1.0 Hz. The optimal precision $\delta\omega^2_\mathrm{opt} = 2e\Gamma/NT$ occurs at $t_\mathrm{opt} = 1/(2N\Gamma) < T$ (red stars) \cite{Huelga1997,Chin2012}. With QZD, $p_0$ exhibits negligible decay. The fitted mean values of $\omega=$19.6 Hz, and $\Gamma$ is remarkably suppressed to 0.1 Hz. The optimal encoding time extends to $t_\mathrm{opt}^\mathrm{Z}=T$, with precision $(\delta\omega^\mathrm{Z}_\mathrm{opt})^2\sim e^{2N\Gamma T}/N^2T^2$, approaching the Heisenberg limit. We also plot the ratio $\delta\omega_\mathrm{opt}/\delta\omega^\mathrm{Z}_\mathrm{opt}$ [green dashed line in Fig.~\ref{Fig3}(b)], thus confirming a $\sqrt{N}$-scaling improvement over the standard approach.

  We further present experimental results in the sequential metrology setting, which exploits time resources to attain high precision with limited physical qubits \cite{Giovannetti2006,juffmann2016,Li2024}. Fixing the qubit number at $N=4$, we extend the encoding time up to $t=2T$.
  As shown in Fig.~\ref{Fig3}(c), in the absence of QZD, $\delta \omega$ rapidly reaches its optimal value of $\delta\omega_\mathrm{opt}=1.71\pm0.09$ Hz, and closely following the theoretical envelope  $\delta\omega = e^{N\Gamma t}/(N\sqrt{2tT})$.
In contrast, with QZD, the precision steadily improves with increasing $t$, in good agreement with the predicted scaling $\delta\omega^\mathrm{Z}=1/N\sqrt{2tT}$.
 The achieved optimal precision, $\delta\omega_\mathrm{opt}=0.56\pm0.03$ Hz, approaches that of parallel setting with $N=8$ and $t=T$ (0.5 Hz, gray dashed line). This demonstrates that our QZD approach can be implemented in the sequential setting, achieving an experimental precision comparable to larger system. To clarify the role of QZD, we perform quantum state tomography at $t=2T$ [Fig.\ref{Fig3}(c)], revealing a final state significantly closer to the GHZ state compared to the no-QZD case, highlighting the protective effect of QZD on the metrological resource.

\textit{Scalability and integration with dynamical decoupling---}In practice, strong interparticle couplings become harder to realize as $N$ increases, making it crucial to understand how the required QZD strength scales with system size. Taking $H_c^{\mathrm{QZ}}$ as an example, we numerically analyze the Ramsey fringe amplitude as a function of the coupling strength $K$ under fixed noise, for various $N$ [Fig.~\ref{Fig4}(a)]. As expected, the amplitude grows with $K$ for a given $N$, signaling the establishment of QZD. Defining $K_\text{th}$ as the smallest $K$ yielding a Ramsey amplitude of 0.9, we find that $K_\text{th}$ scales inversely with $N$, fitting $K_\text{th} = c_1/N+c_2$ (inset). This scaling implies that the required coupling strength decreases with $N$, supporting the scalability of our scheme. In \cite{supp}, we further compare the scalability and robustness of $H_c^{\mathrm{Dip}},H_c^{\mathrm{Is}},H_c^{\mathrm{QZ}}$ for noise suppression.

We also investigate combining QZD with other control techniques to address more general noises in metrology. While QZD suppresses transverse amplitude-damping noise, parallel noise like phase damping often coexists and degrades performance. We show that QZD can be effectively integrated with dynamical decoupling (DD) \cite{Suter2016} to suppress both noise components in ac field frequency estimation. In this setting, the noisy Hamiltonian is $H_\text{en}^\text{ac}+H_\text{AD}+H_\text{PD}$, where $H_\text{en}^\text{ac} = A\sin(\omega t)\sum_i\sigma_z^i/2$, $H_\text{AD} = \xi_{i,x}(t)\sigma_x^i/2$, and $H_\text{PD} = \xi_{i,z}(t)\sigma_z^i/2$. To implement QZD and DD simultaneously, we apply strong continuous coupling $KH_c^\text{Is}$ and introduce periodic $\pi$-pulses along $x$, synchronized with the ac signal's extrema [Fig.\ref{Fig4}(b), inset]. Remarkably, this integrated approach restores the optimal QFI scaling $\mathcal{F}_\omega \sim N^2 T^4$ [Fig.\ref{Fig4}(b)].

 \textit{Summary---}We have proposed a QZD-based scheme to enhance quantum metrology precision in the presence of amplitude noise. Our scheme is successfully verified on a four-qubit NMR processor. Since QZD is implemented by introducing strong inter-qubit couplings, it is well-suited for a broad range of many-body quantum sensors. The combination of QZD with other robust control methods, such as DD, is also investigated numerically, showing promising potential for practical sensing applications. 
 
 The extended simulations in Ref.~\cite{supp} show that within specific noise-frequency ranges, the proposed scheme can effectively resist noises with various noise spectra and those encountered in realistic superconducting quantum systems. Beyond these ranges, stronger couplings are required to retain robustness, which may poses increasing experimental challenges. Our approach is expected to be further integrated with other techniques, such as dynamical control \cite{bai2023,kielinski2024}, quantum error correction \cite{Dur2014,reiter2017}, and nondemolition measurement \cite{Demkowicz2017,Rossi2020}, to recover the quantum advantage in quantum metrology under more complex noise. Owing to its intrinsic ability to preserve coherence, QZD holds promise for broader applications in quantum information processing tasks---such as quantum computation and quantum communication---where maintaining quantum advantages is crucial \cite{Piveteau2022,Graham2022,Conlon2023}.

\begin{acknowledgments}

This work is supported by the National Natural Science Foundation of China (Grants No. 12441502, 12204230, 12404554, 12575020 and 12475042), Innovation Program for Quantum Science and Technology (Grants No. 2024ZD0300400), National Key R\&D Program of China (Grants No. 2023YFF0718400), Guangdong Provincial Quantum Science Strategic Initiative (GDZX2505001, GDZX2303001, GDZX2203001, GDZX2506002, GDZX2403004), Shenzhen Science and Technology Program (Grants No. RCYX20200714114522109).
\end{acknowledgments}

	\bibliographystyle{apsrev4-1}
	
	\providecommand{\noopsort}[1]{}\providecommand{\singleletter}[1]{#1}%

\end{document}